\documentclass[nobibnotes,aps,superscriptaddress,10pt,notitlepage,twocolumn,prl,longbibliography]{revtex4-1}

\usepackage[latin1]{inputenc}
\usepackage{graphicx}
\graphicspath{{Figures/}}
\usepackage{float}
\usepackage{latexsym}
\usepackage{graphicx}
\usepackage{amssymb}
\usepackage{amsmath}
\usepackage{amsfonts}
\usepackage[colorlinks=true,citecolor=blue,linkcolor=blue,hyperfootnotes=false]{hyperref}
\usepackage[dvipsnames]{xcolor}
\usepackage{cool}
\usepackage{dcolumn}
\usepackage{textcomp}
\usepackage{xcolor}
\usepackage{xfrac}
\usepackage{slashed}
\usepackage{multirow}
\usepackage{orcidlink}

\def\be{\begin{equation}}
\def\ee{\end{equation}}
\def\beq{\begin{equation}}
\def\eeq{\end{equation}}
\def\figs/B{B}
\def\bea{\begin{eqnarray}}
\def\eea{\end{eqnarray}}
\def\bg{\begin{eqnarray}}
\def\nd{\end{eqnarray}}

\def\ln{{\rm ln}}

\def\mpl{M_{\rm Pl}}
\def\dd{{\rm d}}
\def\pd{\partial}

\def\calP{{\cal P}}
\def\calR{{\cal R}}

\def\prk{\calP_\calR}

\begin{document}
% ---------------------------------
%%%%%%%%%%%%%%%%%%%%%%%%%%%%%%%%%%%

\title{Resilience of Primordial Black Hole Formation\\from Inflation with a Light Spectator Field}

\author{Dario L.~Lorenzoni \orcidlink{0000-0001-9832-8183}}
\email[Corresponding Author. ]{lorenzod@myumanitoba.ca}
\affiliation{Department of Physics \& Astronomy, University of Manitoba, Winnipeg, MB R3T 2N2, Canada}
\affiliation{Department of Physics, University of Winnipeg, Winnipeg, MB R3B 2E9, Canada}

\author{Sarah R.~Geller \orcidlink{0000-0003-4126-4662}}
\email[Corresponding Author. ]{sarah.geller@yale.edu}
\affiliation{Santa Cruz Institute for Particle Physics, Santa Cruz, CA 95064, USA}
\affiliation{Department of Physics, University of California, Santa Cruz, Santa Cruz, CA 95064, USA}
\affiliation{Physics Division, Lawrence Berkeley National Laboratory, 1 Cyclotron Road, Berkeley, CA, 94720, USA}

\author{Zachary Ireland \orcidlink{0009-0009-8080-6218}}
%\email{zach.ireland@mail.utoronto.ca}
\affiliation{Department of Physics, University of Toronto, Toronto, ON M5S 1A7, Canada}

\author{David I.~Kaiser \orcidlink{0000-0002-5054-6744}}
%\email{dikaiser@mit.edu}
\affiliation{Center for Theoretical Physics---A Leinweber Institute,
Massachusetts Institute of Technology, Cambridge, MA 02139, USA}

\author{Evan McDonough \orcidlink{0000-0002-2130-3903}}
%\email{e.mcdonough@uwinnipeg.ca}
\affiliation{Department of Physics, University of Winnipeg, Winnipeg, MB R3B 2E9, Canada}

\author{Kyle A.~Wittmeier \orcidlink{0009-0008-9914-742X}}
%\email{wittmeier-k@webmail.uwinnipeg.ca}
\affiliation{Department of Physics, University of Winnipeg, Winnipeg, MB R3B 2E9, Canada}

\begin{abstract}
    Scalar fields are ubiquitous in theories of high-energy physics. In the context of cosmic inflation, this suggests the existence of spectator fields, which provide a subdominant source of energy density. We show that spectator fields boost the inflationary production of primordial black holes, with single-field ultra-slow-roll evolution supplanted by a phase of evolution along the spectator direction, and primordial perturbations amplified by the resulting multifield dynamics. This generic mechanism is largely free from the severe fine-tuning that afflicts single-field inflationary PBH models.
\end{abstract}

\maketitle

% =================================
% ---------------------------------
% ---------------------------------
% ---------------------------------
\textbf{\textit{Introduction.}} 
% ---------------------------------
% ---------------------------------
% ---------------------------------
% =================================
Primordial black holes (PBHs) \cite{Zeldovich:1967lct,Hawking:1971ei,Carr:1974nx,Meszaros:1974tb,Carr:1975qj,Khlopov:1985jw,Niemeyer:1999ak}, which form from the direct collapse of overdensities rather than from stellar collapse, are a leading candidate to explain some or all of the mysterious dark matter that fills the Universe ~\cite{Khlopov:2008qy,Carr:2009jm,Sasaki:2018dmp,Carr:2020gox,Carr:2020xqk,Green:2020jor,Escriva:2021aeh,Villanueva-Domingo:2021spv,Escriva:2022duf,Gorton:2024cdm}. 
Perhaps the most well-studied mechanism for PBH formation is the postinflationary collapse of primordial overdensities sourced by curvature fluctuations from inflation. For example, single-field models with a potential $V (\varphi)$ that is tuned to yield brief periods of ultra-slow-roll (USR) evolution are known to produce spikes in the curvature power spectrum \cite{Kinney:2005vj,Martin:2012pe,Ezquiaga:2017fvi,Garcia-Bellido:2017mdw,Germani:2017bcs,Kannike:2017bxn,Motohashi:2017kbs,Di:2017ndc,Ballesteros:2017fsr,Pattison:2017mbe,Passaglia:2018ixg,Biagetti:2018pjj,Mishra:2019pzq,Figueroa:2020jkf,Karam:2022nym,Ozsoy:2023ryl,Cole:2023wyx,Cicoli:2018asa,Cicoli:2022sih,Cai:2022erk,Inomata:2021uqj,Inomata:2021tpx,Bhaumik:2019tvl,Pi:2022ysn,Choudhury:2024one}.
However, demanding these models fit precision cosmic microwave background (CMB) observations \cite{Planck:2019kim,Planck:2018vyg,Planck:2018jri,BICEP:2021xfz} and produce a population of PBHs satisfying all known constraints \cite{Khlopov:2008qy,Carr:2009jm,Sasaki:2018dmp,Carr:2020gox,Carr:2020xqk,Green:2020jor,Escriva:2021aeh,Villanueva-Domingo:2021spv,Escriva:2022duf,Gorton:2024cdm}
requires a significant fine-tuning of model parameters. (Following Ref.~\cite{Cole:2023wyx}, by ``fine-tuning'' we mean that predictions for physical quantities, such as the power spectrum of gauge-invariant curvature perturbations, become exponentially sensitive to small changes in one or more model parameters.) 
On the other hand, multifield inflation models for PBHs have also been proposed (see, e.g., \cite{Randall:1995dj,Garcia-Bellido:1996mdl,Lyth:2010zq,Bugaev:2011wy,Halpern:2014mca,Clesse:2015wea,Kawasaki:2015ppx,Braglia:2022phb,Fumagalli:2020adf,Braglia:2020eai,Palma:2020ejf,Ferraz:2024bvd,Bhattacharya:2022fze,Geller:2022nkr,Qin:2023lgo}), and in certain cases the fine-tuning can be less severe than in the single-field case \cite{Qin:2023lgo}. Multifield models are well motivated by particle physics, e.g., the Standard Model contains four real scalar degrees of freedom, and extensions typically contain many more.

In this paper we identify a broad class of particularly simple multifield models that match CMB observations and yield a population of PBHs. These models include two minimally coupled scalar fields with a potential of the form $V (\varphi, \chi) = V_{\rm PBH} (\varphi) + V_{\rm S} (\chi)$. The inflaton potential $V_{\rm PBH} (\varphi)$ is characterized by a small-field feature, such as a near-inflection point, which leads to a departure from ordinary slow-roll evolution. The field $\chi$ is a light spectator field that does not couple directly to the inflaton, and whose contribution to the total energy density remains subdominant throughout inflation. We consider the simplest potential for the spectator, $V_{\rm S} (\chi) = \frac{1}{2} m_\chi^2 \chi^2$, and study dynamics of the multifield system for various forms of $V_{\rm PBH} (\varphi)$ that lead to PBH formation.  

\begin{figure}[h!]
    \centering
    \includegraphics[width=\linewidth]{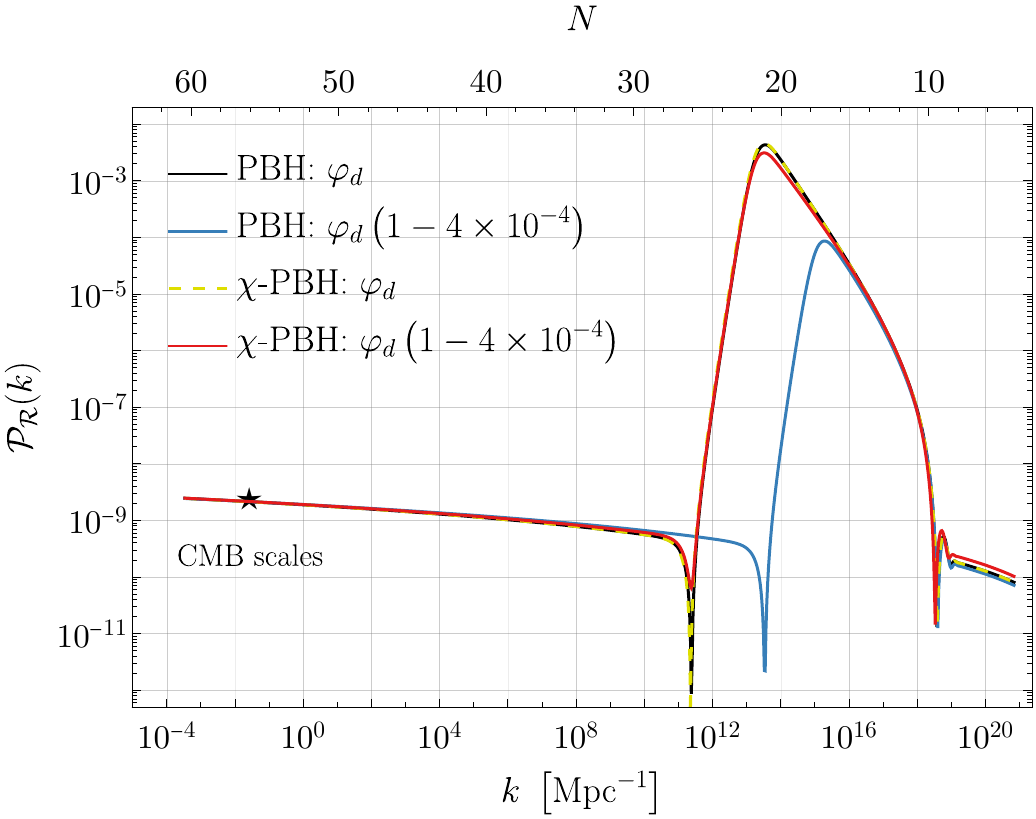}
    \caption{Power spectra ${\cal P}_{\cal R} (k)$ of primordial curvature perturbations 
    for Model $A$ of Eq.~\eqref{eq:VPBH-A}, with and without a spectator field $\chi$. 
     In the single-field case, a small variation of the model parameter $\varphi_d$ eliminates PBH production, shifting the peak of ${\cal P}_{\cal R} (k)$ for fiducial values of the parameters (black curve) below the threshold ($10^{-3}$) required for PBH production (blue curve). (Variations of ${\cal O}(10^{-5})$ in $\varphi_d$ yield ${\cal O}(1)$ shifts in the peak of $\prk(k)$ \cite{Cole:2023wyx}.)  Upon adding the spectator $\chi$, the power spectrum remains resilient to the same small parameter change (yellow-dashed to red curves). Model parameters and corresponding observables are reported in Tables~\ref{tab:PBHA-params} and~\ref{tab:PBHA-obs} of the \hyperref[app:params]{Supplemental Material}.
    }
    \label{fig:PBHA-PRk}
\end{figure}

By adding a simple spectator field to PBH-producing single-field models, we fundamentally change the physical mechanism by which primordial curvature perturbations become amplified. In particular, such models {\it do not undergo a USR phase}. Instead, curvature perturbations on scales $k_{\rm PBH} \gg k_{\rm CMB}$ become amplified by {\it tachyonic growth} of {\it isocurvature perturbations} combined with {\it turns in field space}---features that have no analog in single-field models. This provides a generic mechanism~\footnote{Previous studies have highlighted model-dependent scenarios in which spectator fields can enhance primordial curvature perturbations, and hence PBH formation \cite{Stamou:2023vxu,Stamou:2024xkk,Stamou:2024lqf,Wilkins_2024,Kuroda:2025coa,Kohri:2012yw}. } that is applicable to any base model in which $V_{\rm PBH} (\varphi)$ (when considered in a single-field context) yields a spike in the primordial curvature power spectrum. The mechanism exhibits  a {\it resilience} that is absent in the corresponding single-field cases: the models can match CMB observations and produce an appropriate population of PBHs while alleviating the exponential sensitivity to small changes in parameters that plagues the single-field models. See Fig.~\ref{fig:PBHA-PRk}.\\

% =================================
% ---------------------------------
% ---------------------------------
% ---------------------------------
\textbf{\textit{Inflationary Dynamics.}}
% ---------------------------------
% ---------------------------------
% ---------------------------------
% =================================
We consider two-field models governed by the action
\begin{equation}\label{eq:action}
    S= \int \dd^4 x \sqrt{ -g} \left[ \frac{M_\text{Pl}^2}{2} R - \frac{1}{2} (\partial \varphi)^2 - \frac{1}{2}(\partial \chi)^2 - V(\varphi,\chi)\right]\,,
\end{equation}
where $M_{\text{Pl}}\equiv1/\sqrt{8\pi G}$ is the reduced Planck mass and $V(\varphi,\chi) = V_{\rm PBH}(\varphi) + V_{\rm S}(\chi)$,  with $V_{\rm S} (\chi) = \frac{1}{2} m_\chi^2 \chi^2$. \footnote{We define a scalar spectator field as being subdominant in potential energy density, $\left|V_\text{\rm S}/V_\text{PBH}\right|_{t_{\text{CMB}}} \ll 1$, with no direct coupling to the inflaton, and with a sub-Hubble mass  $\left|m_\chi/H\right|_{t_{\text{CMB}}} \ll 1$, where $H$ is the Hubble parameter. Both conditions are evaluated at the time $t_{\text{CMB}}$ when the comoving CMB pivot scale $k_{\rm CMB}$ first exits the Hubble radius.  }
The evolution of the background fields is governed by their equations of motion and the Friedmann equation,
\begin{gather}
    \ddot{\varphi}+3 H \dot{\varphi} = - \frac{\dd V_{\rm PBH}}{\dd \varphi} \, ,\quad 
    \ddot{\chi}+3 H \dot{\chi} = - \frac{\dd V_{\rm S}}{\dd \chi}\,, 
    \label{eq:EoM-bckgr}
    \\
    H^2=\frac{1}{3M_\text{Pl}^2}\left( \frac{\dot{\varphi}^2}{2} + \frac{\dot{\chi}^2}{2} + V_{\rm PBH} + V_\text{S}\right)\,.
    \label{eq:Friedmann}
\end{gather}
Note that the inflaton field $\varphi$ and the spectator $\chi$ are coupled only via the Friedmann equation, unlike the interesting multifield models studied in Refs.~\cite{Fumagalli:2020adf,Braglia:2020eai,Palma:2020ejf,Geller:2022nkr,Qin:2023lgo}. In other words, the dynamics we identify here do not require a curved field-space manifold with its associated direct coupling of the fields via a noncanonical kinetic sector, and neither is the amplification of the primordial power spectrum triggered by having more than one phase of inflation (as in Ref.~\cite{Braglia:2020eai}). In the present cases,
$\chi$ contributes a small increase to $H$ and hence to the Hubble friction
term for $\varphi$.
For suitable choice of potentials, these equations admit inflationary solutions characterized by the slow-roll parameters $\epsilon \equiv - \dot{H}/{H^2}$ and $\eta \equiv 2 \epsilon - \dot{\epsilon}/(2H \epsilon)$. The decoupling of the fields implies that the first slow-roll parameter is sum separable,
\begin{equation}\label{eq:epsilon}
    \epsilon
    =\frac{1}{2}\frac{\dot{\varphi}^2}{H^2 M_\text{Pl}^2} + \frac{1}{2}\frac{\dot{\chi}^2}{H^2 M_\text{Pl}^2} \equiv \epsilon_\varphi+\epsilon_\chi\,.
\end{equation}
To study the dynamics of the two-field system we deploy the usual formalism for multifield models \cite{Sasaki:1995aw,Gordon:2000hv,Wands:2002bn,Langlois:2008mn,Peterson:2010np,Gong:2011uw,Kaiser:2012ak,Gong:2016qmq}.
Using the notation $\phi^I\equiv \{\varphi,\chi\}$, we may define the magnitude of the velocity of the background fields as $\dot{\sigma}\equiv|\dot{\phi}^{I}| = \sqrt{\dot{\varphi}^2 + \dot{\chi}^2}$.
We can then define a field-space unit vector $\hat{\sigma}^{I} \equiv \dot{\phi}^I / \dot{\sigma} = \epsilon^{-1/2} \{ \sqrt{ \epsilon_\varphi} , \sqrt{ \epsilon_\chi} \}$ that points along the direction of the background fields' motion (the \textit{adiabatic} direction). We also define the vector $\hat{s}^J \equiv \varepsilon^{IJ} \hat{\sigma}_I$ (the \textit{isocurvature} direction) which is orthogonal to $\hat{\sigma}^I$. (Here
$\varepsilon^{IJ}$ is the two-dimensional Levi-Civita antisymmetric pseudotensor.) 
The time evolution of these unit vectors is captured by the turn rate pseudovector and pseudoscalar, 
\begin{equation}\label{eq:turn-rate}
    \omega^I\equiv\pd_t\hat{\sigma}^I 
    \,,\quad
    \omega\equiv \varepsilon_{IJ}\hat{\sigma}^I\omega^J\,.
\end{equation}
In the simple two-field models studied here, the turn rate is $\omega=\dot{\theta}$, where $\theta$ is the angle $\hat{\sigma}^I$ makes with the $\varphi$ axis.

We may also consider the gauge-invariant fluctuations along the adiabatic (or \textit{curvature}) and isocurvature directions, 
and decompose them into Fourier modes denoted ${\cal R}_{k}$ and ${\cal S}_k$, respectively \cite{Wands:2007bd,Gong:2016qmq}. (We consider perturbations around a spatially flat Friedmann-Lema\^{i}tre-Robertson-Walker line-element.) To linear order, these obey the coupled equations of motion \cite{Kaiser:2012ak,Achucarro:2016fby,McDonough:2020gmn,Lorenzoni:2024krn}
\begin{gather}
    \label{eq:EoM-R}
    \frac{d}{dt}  \left(\dot {\cal R}_k - 2 \omega {\cal S}_k \right)
    + (3 + \delta) H \left( \dot  {\cal R}_k - 2 \omega {\cal S}_k \right)  + \frac{k^2}{a^2}{\cal R}_k = 0\,,\\
    \label{eq:EoM-S}
    \ddot{\cal S}_k+ (3+ \delta) H \dot{\cal S}_k + \left(\frac{k^2}{a^2} + \mu^2_s\right){\cal S}_k = - 2 \omega \dot{\cal R}_k\,.
\end{gather}
Here $\delta\equiv \dot{\epsilon}/(H\epsilon)= 4\epsilon-2\eta$ and $\mu_s$ is the  
isocurvature mass \cite{Kaiser:2012ak,Achucarro:2016fby,McDonough:2020gmn,Lorenzoni:2024krn}
\begin{equation}\label{eq:muS}
    \mu_s ^2 = {\cal M}_{ss} - {\cal M}_{\sigma\sigma} + 2 H^2 \epsilon\left(3+\delta-\epsilon\right)\, ,
\end{equation}
where ${\cal M}_{\sigma\sigma}$ and ${\cal M}_{ss}$ are projections of the second derivatives of $V (\varphi, \chi)$
onto the adiabatic and isocurvature directions \cite{notemus}. The dimensionless power spectra for modes ${\cal X} \in \{ {\cal R}, {\cal S} \}$ are defined as ${\cal P}_{ {\cal X}} (k) \equiv k^3 \vert {\cal X}_k \vert^2 / (2 \pi^2)$.
On super-Hubble scales ($k/a \ll H$), the system vastly simplifies: the curvature and isocurvature modes satisfy 
\begin{gather}
    \label{eq:EoM-S-superHubble}
    \dot{\cal R}_k \simeq 2 \omega {\cal S}_k \,, \quad
    \ddot{\cal S}_k+ (3+ \delta) H \dot{\cal S}_k + \tilde{\mu}^2_s {\cal S}_k \simeq 0\,,
\end{gather}
where the effective isocurvature mass (in the long-wavelength limit) is $\tilde{\mu}_s^2\equiv\mu_s^2 + 4\omega^2$. The isocurvature modes ${\cal S}_k$ will experience tachyonic growth when $\tilde{\mu}_s^2<0$ and will decay when $\tilde{\mu}_s^2>0$. \\

% =================================
% ---------------------------------
% ---------------------------------
% ---------------------------------
\textbf{\textit{Single-Field Cases.}}
% ---------------------------------
% ---------------------------------
% ---------------------------------
% =================================
In single-field models, for which ${\cal S}_k = 0$ and $\omega = 0$ identically, the gauge-invariant curvature perturbations ${\cal R}_k$ will become amplified on certain scales $k_{\rm PBH}$ because of changes in the evolution of the background field $\varphi$. In particular, note from Eq.~(\ref{eq:EoM-R}) that the Hubble damping term, $(3 + \delta )H \dot{\cal R}_k$, will change sign whenever $(3 + \delta) = (3 + 4 \epsilon - 2 \eta)$ becomes negative. This is exactly what happens during USR, when $\epsilon \rightarrow 0^+$ and $\eta \rightarrow 3$. Then long-wavelength modes, with $k^2 / (a H)^2 < \vert \dot{\cal R}_k / ( H {\cal R}_k) \vert$, will become {\it antidamped} and grow quasiexponentially whenever $\epsilon \ll 1$ and $\eta \geq 3/2$ \cite{Kinney:2005vj,Martin:2012pe,Kannike:2017bxn,Germani:2017bcs,Ezquiaga:2017fvi,Garcia-Bellido:2017mdw,Geller:2022nkr,Qin:2023lgo}.

Single-field models will enter a USR phase when the derivative of the potential becomes nearly flat, $\partial_\varphi V_{\rm PBH} (\varphi) \sim 0$, in the vicinity of which $\varphi (t)$ will evolve as $\ddot{\varphi} \simeq - 3 H \dot{\varphi}$ \cite{Kinney:2005vj,Martin:2012pe,Geller:2022nkr,Qin:2023lgo}. Whereas it is usually straightforward to arrange the functional form of $V_{\rm PBH} (\varphi)$ to include some portion for which $\partial_\varphi V_{\rm PBH} (\varphi) \sim 0$, requiring that the system does not {\it remain} in USR arbitrarily long while also ensuring that the small-field feature within $V_{\rm PBH} (\varphi)$ does not distort predictions for CMB observables on scales $k_{\rm CMB}$ necessarily entails significant fine-tuning \cite{Cole:2023wyx}. 

For concreteness, we consider two different forms of $V_{\rm PBH} (\varphi)$ which have been well studied in the literature:
\begin{gather}
    V_{{\rm PBH},A} (\varphi) = V_0 \frac{ \varphi^2}{\varphi^2 + M^2} \left( 1 + A e^{- (\varphi - \varphi_d)^2/2\sigma^2} \right) , \label{eq:VPBH-A}\\
    V_{{\rm PBH},B} (\varphi) = \frac{ M_{\rm Pl}^4 \lambda v^4}{12} \frac{ x^2 \left( 6 - 4 a x + 3 x^2 \right)}{\left( M_{\rm Pl}^2 + b x^2 \right)^2} ,  \label{eq:VPBH-B}
\end{gather}
with $x \equiv \varphi / v$. Model $A$ \cite{Mishra:2019pzq} superimposes a local Gaussian ``bump'' on the well-known Kachru-Kallosh-Linde-Trivedi (KKLT) potential \cite{Kachru:2003aw}. Model $B$ \cite{Garcia-Bellido:2017mdw,Germani:2017bcs} adds a cubic term to a potential akin to Higgs inflation \cite{Bezrukov:2007ep} to engineer a quasi-inflection point. Consistent with Ref.~\cite{Cole:2023wyx}, we find that small variations of model parameters, of fractional size ${\cal O}(10^{-5})$, alter the peak amplitude of the power spectrum by a fractional change of ${\cal O}(1)$. In cases for which the peak of the original power spectrum was above the threshold to produce PBHs but still within the perturbative regime, with ${\cal O}(1) \times 10^{-3} < {\cal P}_{\cal R}(k_{\rm peak}) \ll 1$, such an ${\cal O}(1)$ fractional change to ${\cal P}_{\cal R} (k_{\rm peak})$ generically will not shut off PBH formation. On the other hand, fractional parameter variations of ${\cal O} (10^{-3})$ sufficiently alter the growth of modes ${\cal R}_k$ so as to yield an ${\cal O} (10)$ fractional change in ${\cal P}_{\cal R} (k_{\rm peak})$, pushing the power spectrum below the threshold required to produce PBHs. See Figs.~\ref{fig:PBHA-PRk} and \ref{fig:PBHB-PRk} and Tables~\ref{tab:PBHA-params} and~\ref{tab:PBHA-obs} in the \hyperref[app:params]{Supplemental Material}.\\

\begin{figure}[h!]
    \centering
    \includegraphics[width=\linewidth]{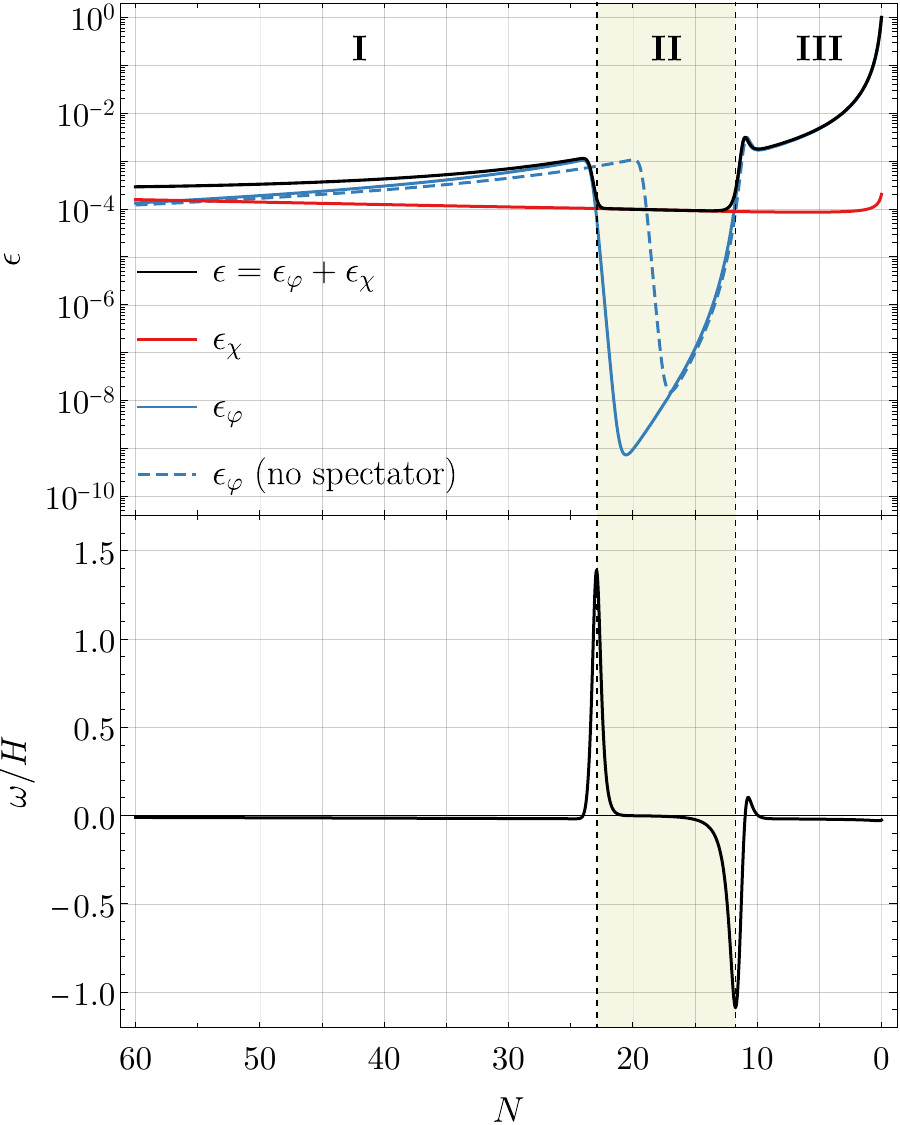}
    \caption{
    Evolution of $\epsilon$ and $\omega$ as functions of the number of efolds $N$ before the end of inflation, for $V = V_{{\rm PBH},A} (\varphi) + V_{\rm S} (\chi)$ of Eq.~\eqref{eq:VPBH-A}. 
    \emph{Top panel:} Slow-roll parameter $\epsilon$ and its components; for comparison, the evolution of $\epsilon_\varphi$ for the spectator-less case is also shown.
    \emph{Bottom panel:} Turn rate pseudoscalar $\omega/H$, showing the change in trajectory of the background field system; the dashed vertical lines correspond to the extrema of $\omega$.
    \emph{Both panels:} The highlighted ``phase II'' is delimited by the turns;
    during this time, $\epsilon_\chi\gg \epsilon_\varphi$. Model parameters are reported in Table~\ref{tab:PBHA-params}.
   }
    \label{fig:PBHA-background}
\end{figure}

% =================================
% ---------------------------------
% ---------------------------------
% ---------------------------------
\textbf{\textit{Role of the Spectator Field.}} 
% ---------------------------------
% ---------------------------------
% ---------------------------------
% =================================
Figure~\ref{fig:PBHA-background} shows $\epsilon$ and $\omega$ for the system $V (\varphi, \chi) = V_{{\rm PBH},A} (\varphi) + V_{\rm S} (\chi)$. Much as in the single-field version of this model (setting $\chi = 0$), the slow-roll parameter $\epsilon_\varphi$ becomes anomalously small, $\epsilon_\varphi \sim {\cal O} (10^{-9})$, as $\varphi$ encounters the region of its potential in which $\partial_\varphi V_{ {\rm PBH}, A} (\varphi) \sim 0$ (solid and dashed blue curves of Fig.~\ref{fig:PBHA-background}, top panel). Yet in the two-field case, since $V_{\rm S} (\chi)$ has no special features, $\chi$ never departs from ordinary slow-roll, with $\epsilon_\chi \sim {\cal O} (10^{-4})$ throughout inflation (red curve). Hence $\epsilon \geq \epsilon_\chi$ never becomes anomalously small, and the system never enters USR. (We have also confirmed that $\eta < 3$ throughout inflation.) Meanwhile, $\chi$ backreacts on $\varphi$ via its contribution to $H$, increasing Hubble friction and lengthening the duration of phase II compared to the USR phase in the single-field case. 

The fact that $\epsilon_\chi$ never departs from ordinary slow-roll forces the system to undergo two turns. Recall that $\hat{\sigma}^I = \epsilon^{-1/2} \{ \sqrt{ \epsilon_\varphi} , \sqrt{ \epsilon_\chi} \}$; the direction of the system's evolution through field space depends on the ratio of $\epsilon_\varphi$ to $\epsilon_\chi$ over time. During phase I, $\epsilon_\varphi > \epsilon_\chi$, and the system predominantly evolves along the $\varphi$ direction. Once $\epsilon_\varphi$ begins to fall exponentially, suddenly $\epsilon_\chi \gg \epsilon_\varphi$, and the unit vector $\hat{\sigma}^I$ rapidly turns to point along the $\chi$ direction. By construction, the inflaton $\varphi$ must eventually ``escape'' the flat region and reach the global minimum of $V_{ {\rm PBH}, A} (\varphi)$; this requires that $\varphi$ roll down from a local maximum, at which $\partial_\varphi^2 V_{ {\rm PBH}, A} (\varphi) < 0$. As $\varphi$ accelerates, $\epsilon_\varphi$ grows and eventually exceeds $\epsilon_\chi$, causing a second sharp turn into phase III.

This pattern---two turns bracketing a phase II during which $\epsilon_\chi \gg \epsilon_\varphi$---is {\it generic} for models of the form $V (\varphi, \chi) = V_{{\rm PBH}} (\varphi) + V_{\rm S} (\chi)$, in which $V_{\rm PBH} (\varphi)$ includes features that drive a USR phase in the single-field case and $\chi$ is a light spectator field. Regardless of the particular form that $V_{\rm PBH} (\varphi)$ takes, it must include some region with $\partial_\varphi V_{\rm PBH} (\varphi) \sim 0$ and a neighboring region in which $\partial_\varphi^2 V_{\rm PBH} (\varphi) < 0$, in order for $\varphi$ to enter and later exit a USR phase. For example, Fig.~\ref{fig:PBHB-background} shows the same three-phase structure for $V(\varphi, \chi) = V_{{ \rm PBH},B} (\varphi) + V_{\rm S} (\chi)$, even though $V_{{\rm PBH},A} (\varphi)$ and $V_{{\rm PBH},B} (\varphi)$ in Eqs.~(\ref{eq:VPBH-A})--(\ref{eq:VPBH-B}) differ considerably in functional form.

\begin{figure}[h!]
    \centering
    \includegraphics[width=\linewidth]{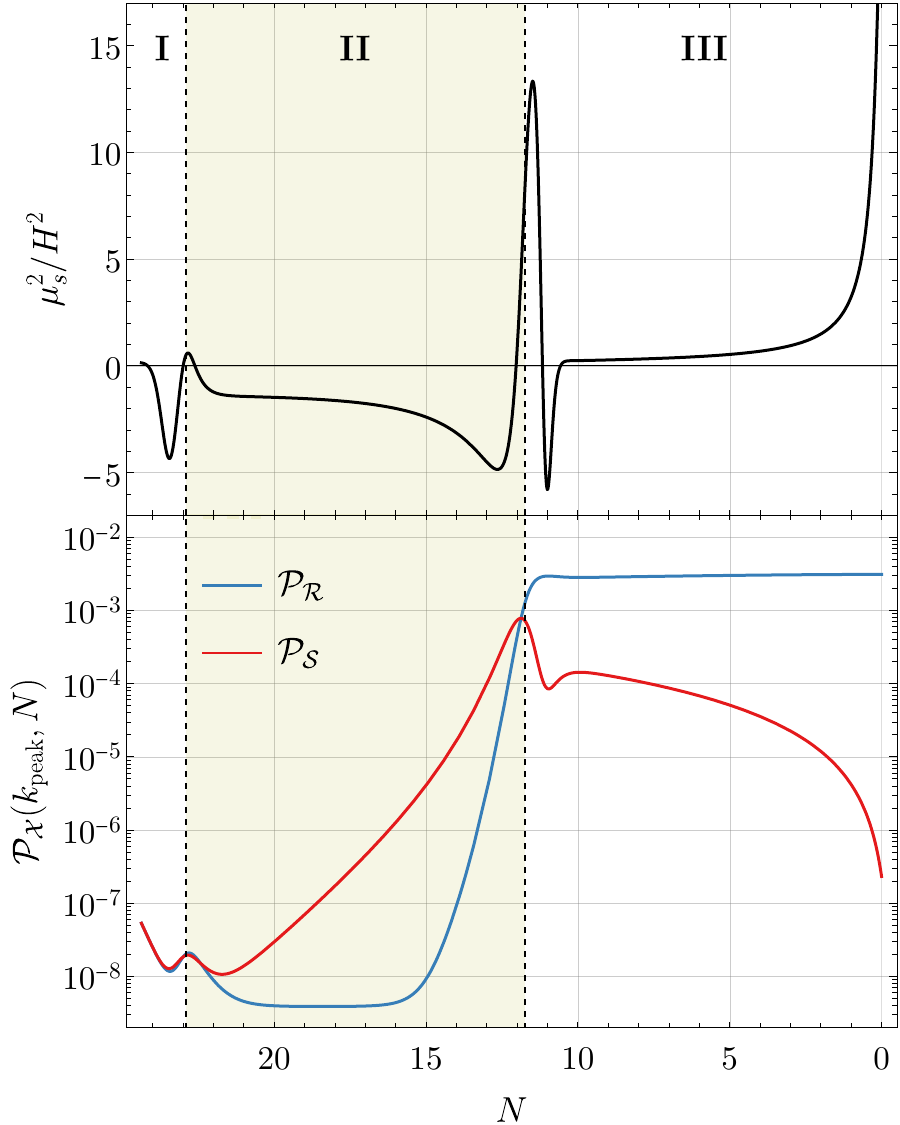}
    \caption{Evolution of the perturbations as functions of the number of e-folds $N$ before the end of inflation, for $V = V_{{\rm PBH},A} (\varphi) + V_{\rm S} (\chi)$ of Eq.~(\ref{eq:VPBH-A}).
    \textit{Top panel:} Isocurvature effective mass $\tilde{\mu}_s^2$; the tachyonic instability is highlighted in phase II. 
    \textit{Bottom panel:} Spectra of curvature (${\cal R}$) and isocurvature (${\cal S}$) perturbations for a mode $k_\text{peak}$ that exits the Hubble radius $\sim 23$ e-folds before the end of inflation. The mode ${\cal S}_{k_{\rm peak}}$ is amplified exponentially via tachyonic growth while $\tilde{\mu}_s^2 < 0$, and transfers that power to ${\cal R}_{k_{\rm peak}}$ at the second turn, which leads to phase III. Model parameters (``$\chi$-PBH$A$-var'') are listed in Table~\ref{tab:PBHA-params}. 
    }
    \label{fig:PBHA-perturbations}
\end{figure}

The three-phase pattern of background dynamics has dramatic consequences for the evolution of the perturbations ${\cal R}_k$ and ${\cal S}_k$. As shown in Eqs.~(\ref{eq:EoM-R})--(\ref{eq:EoM-S}), the curvature and isocurvature modes couple only when $\omega \neq 0$. Moreover, for the models in the class we consider here, during phase II the isocurvature mass $\mu_s^2$ generically becomes tachyonic. Following the first turn, when $\epsilon_\chi \gg \epsilon_\varphi$, the adiabatic direction (temporarily) aligns along the $\chi$ direction, and hence the isocurvature direction points in the $\varphi$ direction. During phase II, then, every term within Eq.~(\ref{eq:muS}) becomes negative: ${\cal M}_{ss} \rightarrow \partial_\varphi^2 V_{\rm PBH} (\varphi)$; ${\cal M}_{\sigma\sigma} \rightarrow \partial_\chi^2 V_{\rm S} (\chi)$; and $(3 + \delta - \epsilon) \rightarrow (3 - 2 \eta) < 0$ for $\eta > 3/2$. As described above, during phase II the inflaton potential {\it must} become concave,
with $\partial_\varphi^2 V_{\rm PBH} (\varphi) < 0$; and $\partial_\chi^2 V_{\rm S} (\chi) = m_\chi^2 > 0$ always. Hence---aside from the brief periods surrounding each sharp turn, when $\omega^2 \gg H^2$---both $\mu_s^2 < 0$ and $\tilde{\mu}_s^2 < 0$ for most of phase II. (See Fig.~\ref{fig:PBHA-perturbations}, top panel.) 

This phase of tachyonic instability drives rapid growth in the isocurvature modes ${\cal S}_k$. When the system encounters the second turn (onset of phase III), the now-large modes ${\cal S}_k$ transfer their power to curvature perturbations ${\cal R}_k$ and rapidly decay (since $\tilde{\mu}_s^2 > 0$ in phase III), yielding an exponential increase of ${\cal P}_{\cal R} (k_{\rm PBH})$ for comoving wavenumbers $k_{\rm PBH}$ that were subject to the tachyonic growth (see Fig.~\ref{fig:PBHA-perturbations}, bottom panel)~\footnote{By adding a light spectator field to single-field models, the PBH-forming dynamics more closely resemble the well-studied physics of hybrid inflation \cite{Randall:1995dj,Garcia-Bellido:1996mdl,Lyth:2010zq,Bugaev:2011wy,Halpern:2014mca,Clesse:2015wea,Kawasaki:2015ppx,Braglia:2022phb,Fumagalli:2020adf}: a sharp turn in field space combined with a brief phase of tachyonic growth seeds PBH formation.}.\\

% =================================
% ---------------------------------
% ---------------------------------
% ---------------------------------
\textbf{\textit{Sensitivity to small parameter changes.}}
% ---------------------------------
% ---------------------------------
% ---------------------------------
% =================================
Single-field models that produce PBHs via a phase of USR require substantial fine-tuning if they are also to maintain a close match to CMB observations \cite{Cole:2023wyx}. This is clear in Figs.~\ref{fig:PBHA-PRk} for $V_{{\rm PBH},A} (\varphi)$ and \ref{fig:PBHB-PRk} for $V_{{\rm PBH},B} (\varphi)$. In each case, a selection of fiducial parameters (black curves) yields power spectra ${\cal P}_{\cal R} (k)$ that are in close agreement with CMB measurements and also yield PBHs with masses in the so-called asteroid-mass range, $10^{17} \,{\rm g} \leq M_{\rm PBH} \leq 10^{23} \, {\rm g}$, which would be viable candidates for most or all of dark matter \cite{Khlopov:2008qy,Carr:2009jm,Sasaki:2018dmp,Carr:2020gox,Carr:2020xqk,Green:2020jor,Escriva:2021aeh,Villanueva-Domingo:2021spv,Escriva:2022duf,Gorton:2024cdm}. (See Tables~\ref{tab:PBHA-obs} and~\ref{tab:PBHB-obs}.) However, a tiny variation of ${\cal O}(10^{-3})$ in $\varphi_d$ (for Model $A$) or $v$ (for Model $B$) changes the location and peak of ${\cal P}_{\cal R} (k)$ considerably (blue curves)---such that, in each case, ${\cal P}_{\cal R} (k) \ll 10^{-3}$, remaining well below the threshold required to induce gravitational collapse and produce PBHs \cite{Young:2019yug,Kehagias:2019eil,Escriva:2019phb,DeLuca:2020ioi,Musco:2020jjb,Escriva:2021aeh}.

\begin{figure}[h!]
    \centering
    \includegraphics[width=\linewidth]{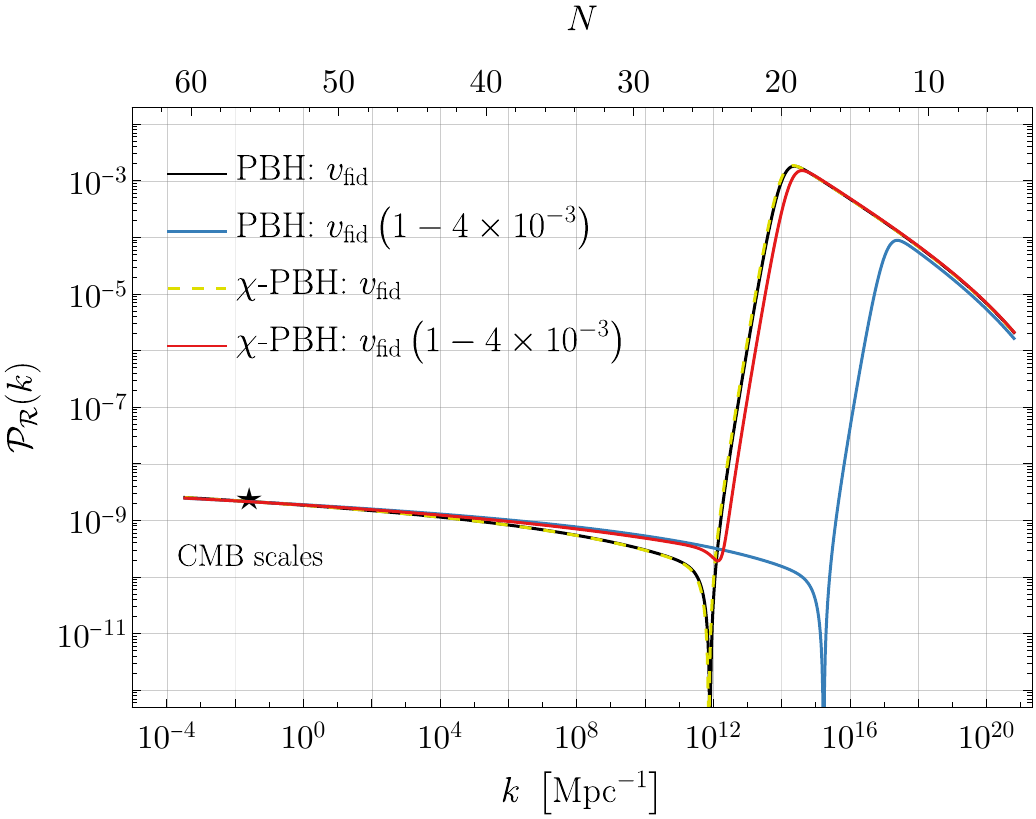}
    \caption{Power spectra ${\cal P}_{\cal R} (k)$ of primordial curvature perturbations for Model $B$ of Eq.~\eqref{eq:VPBH-B}, with and without a spectator field $\chi$. 
    Model parameters and corresponding observables are reported in Tables~\ref{tab:PBHB-params} and~\ref{tab:PBHB-obs}. 
    }
    \label{fig:PBHB-PRk}
\end{figure}

Including a simple light spectator $\chi$ relaxes this problem. For each model, we first find non-fine-tuned values for the spectator parameters $m_\chi$ and $\chi_i$ that reproduce the original single-field spectra with fiducial values of parameters in $V_{\rm PBH}(\varphi)$ (yellow-dashed curves in Figs.~\ref{fig:PBHA-PRk} and \ref{fig:PBHB-PRk}). We then find that a change of ${\cal O} (10^{-3})$ in a fiducial parameter of $V_{\rm PBH} (\varphi)$ is compensated by a coarse-grained ${\cal O} (1)$ variation in $m_\chi$ and $\chi_i$ (red curves in Figs.~\ref{fig:PBHA-PRk} and \ref{fig:PBHB-PRk}). In each case involving the spectator field, predictions for CMB observables and for PBHs
relevant for dark matter remain in strong agreement with CMB measurements~\cite{Planck:2019kim,Planck:2018vyg,Planck:2018jri,BICEP:2021xfz} and PBH constraints~\cite{Khlopov:2008qy,Carr:2009jm,Sasaki:2018dmp,Carr:2020gox,Carr:2020xqk,Green:2020jor,Escriva:2021aeh,Villanueva-Domingo:2021spv,Escriva:2022duf,Gorton:2024cdm}. This is accomplished without overfitting: models such as $V_{{\rm PBH}, B}(\varphi) + V_{\rm S} (\chi)$ match eight observables $(A_s, n_s, \alpha_s, r, \beta_{\rm iso}, f_{\rm NL}, {\cal P}_{\cal R} (k_{\rm peak}), M_{\rm PBH})$---each defined in the Supplemental Material---using only six free parameters $(\lambda, v, a, b, m_\chi, \chi_i)$. 

We may easily estimate the range of spectator parameters that are relevant for the mechanism.
We require $\epsilon_\chi>\epsilon_\varphi$ during phase II. From Eqs.~\eqref{eq:EoM-bckgr} and~\eqref{eq:epsilon}, we find $\epsilon_\chi = \frac{1}{18} (\chi_i / \mpl)^2 (m_\chi / H)^4$.
Taking $\epsilon_\varphi\sim{\cal O}(10^{-10})$ during USR and considering $\chi_i$ comparable to $\varphi_i$, we estimate $10^{-3}\lesssim m_\chi/H\ll 1$, where the upper bound is required for $\chi$ to be a light spectator.\\

% =================================
% ---------------------------------
% ---------------------------------
% ---------------------------------
\textbf{\textit{Discussion.}}
% ---------------------------------
% ---------------------------------
% ---------------------------------
% =================================
Spectator fields are a generic prediction of high-energy physics, arising already in the context of the Standard Model. In this Letter we have shown that in simple scenarios that include at least one light spectator field in addition to an inflaton exhibiting a small-field feature, multifield dynamics will {\it generically} amplify modes ${\cal R}_k$. The background fields will undergo {\it turns} in field space as the system evolves toward the global minimum of the potential. Between those turns, certain modes $k$ of the gauge-invariant isocurvature perturbations ${\cal S}_k$ will grow via {\it tachyonic instability}, and then {\it transfer power} to the curvature perturbations ${\cal R}_k$ before the end of inflation. Simple models involving at least one spectator field can match all CMB observables to high precision and produce a population of PBHs relevant for dark matter, while remaining resilient to small changes in model parameters. 

This family of models can lead to additional phenomenology, notably due to a possible second $\chi$-dominated phase of inflation \textit{\`a la} double inflation~\cite{Silk:1986vc,Salopek:1988qh,Polarski:1992dq} taking place after the decay of $\varphi$. 
This can be induced using a large value $\chi_i$, and could shift the peak of ${\cal P}_{\cal R} (k)$ 
to larger scales, possibly sourcing the production of PBH seeds for the supermassive black holes found at the centers of galaxies~\cite{Clesse:2015wea,Nakama:2016kfq,Carr:2018rid,Cai:2023ptf}.
In this work, spectator parameters were chosen to avoid this second phase, to facilitate comparison with the original phenomenology. 
We leave this analysis, together with the study of reheating dynamics, for future work.

Finally, we note that since the system of inflaton plus spectator never enters USR---and hence neither slow-roll parameter ever becomes anomalously small---it is likely that such two-field models avoid any dangerous growth of loop corrections that some have argued could threaten perturbative control for single-field USR models. (Compare Refs.~\cite{Cheng:2021lif,Kristiano_yokoyama_1,Kristiano_yokoyama_2,sayantan_1,sayantan_2,Cheng:2023ikq} with ~\cite{Senatore_2010,Senatore:2012nq,Pimentel:2012tw,Senatore:2012ya,Ando:2020fjm,RiottoPBH,Firouzjahi:2023aum,Motohashi:2023syh,Firouzjahi:2023ahg,Franciolini:2023agm,Tasinato:2023ukp}.) Possible roles of loop corrections or other nonperturbative effects \cite{Caravano:2024tlp,Caravano:2024moy,Inomata:2022yte,Braglia:2025cee,Braglia:2025qrb,Choudhury:2023rks} remain the subject for further research.\\

% =================================
% ---------------------------------
% ---------------------------------
% ---------------------------------
\textbf{\textit{Acknowledgements.}}
% ---------------------------------
% ---------------------------------
% ---------------------------------
% =================================
The authors thank Josu Aurrekoetxea, Matteo Braglia, Bryce Cyr, Alexandra Klipfel, Jerome Martin, Sebastien Renaux-Petel, Vincent Vennin, and Rainer Weiss for helpful discussions. Portions of this research were conducted in MIT's Center for Theoretical Physics -- a Leinweber Institute and supported in part by the U.S.~Department of Energy under Contract No.~DE-SC0012567. E.M. is supported in part by a Discovery Grant from the Natural Sciences and Engineering Research Council of Canada, and by a New Investigator Operating Grant from Research Manitoba. S.R.G. is supported by the NSF Mathematical and Physical Sciences Ascending postdoctoral fellowship, under Award No.~2317018.\\

\vskip -15pt

\bibliography{refs-spectators}

\clearpage
\appendix
\setcounter{equation}{0}
\setcounter{table}{0}
\setcounter{figure}{0}
\renewcommand{\theequation}{SM\arabic{equation}}
\renewcommand{\thetable}{SM\Roman{table}}
\renewcommand{\thefigure}{SM\arabic{figure}}

% =================================
% ---------------------------------
% ---------------------------------
% ---------------------------------
\section*{Supplemental Material:\\Model Parameters and Observables}\label{app:params}
% ---------------------------------
% ---------------------------------
% ---------------------------------
% =================================

We have performed all calculations of both background quantities and perturbations using two independent codes written in Python and Mathematica, respectively, and find excellent agreement between the two. In Python, we use the publicly available code \texttt{PyTransport}~\cite{Dias:2016rjq,Mulryne:2016mzv}, which uses the Transport method \cite{Dias:2015rca} to directly compute the time-evolution of correlation functions. In Mathematica, we solve the perturbation equations in the form presented in Ref.~\cite{Achucarro:2010da} and, from this, construct the power spectra. Versions of each of these codes are available at the public GitHub repository \texttt{PBHspec}~\cite{CODEnote}.

\begin{figure}[h]
    \centering
    \includegraphics[width=\linewidth]{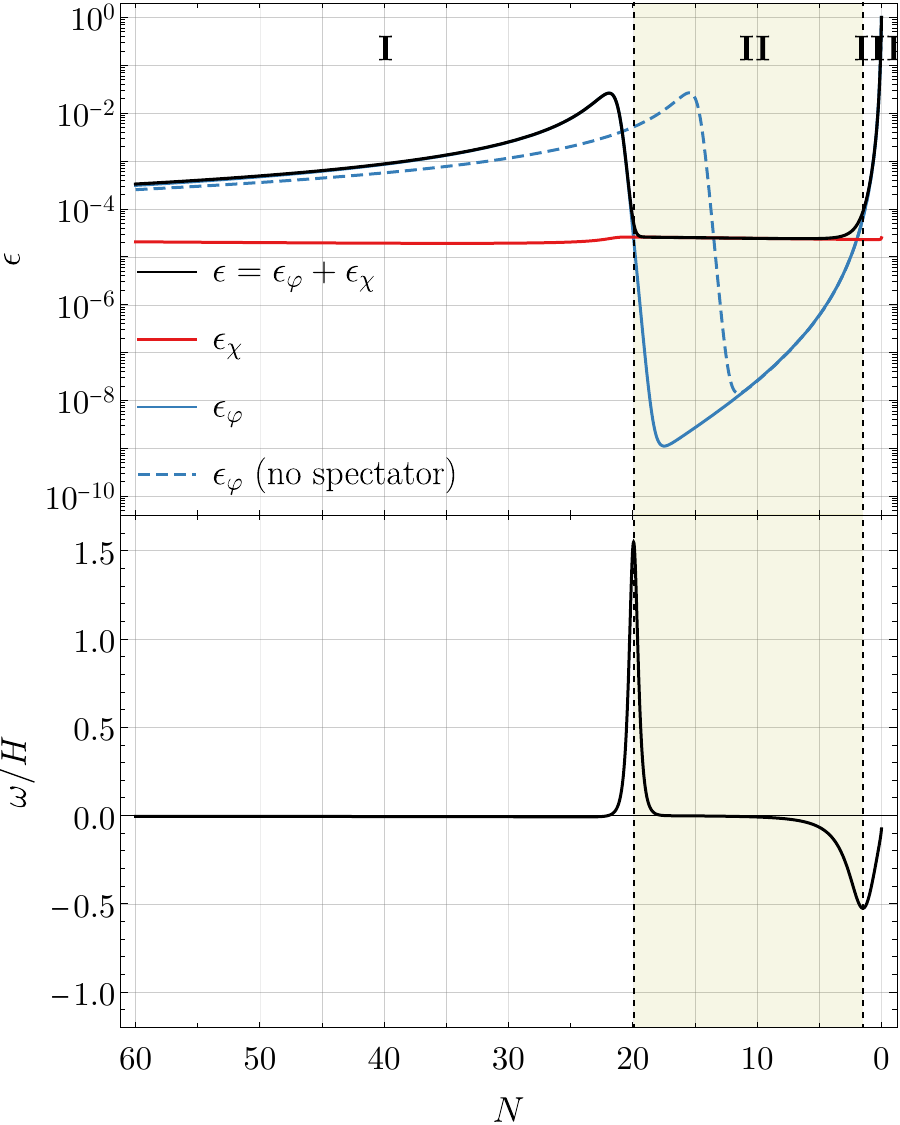}
    \caption{
    Evolution of the background quantities as functions of the number of efolds $N$ before the end of inflation, for the PBH$B$ model of Eq.~\eqref{eq:VPBH-B}, with parameters reported in Table~\ref{tab:PBHB-params}.
    \emph{Top panel:} Slow-roll parameter $\epsilon$ and its components; for comparison, the evolution of $\epsilon$ for the spectator-less case is also shown.
    \emph{Bottom panel:} Turn rate pseudoscalar $\omega/H$, showing the change in trajectory of the background field system; the dashed vertical lines correspond to the extrema of $\omega$.
    \emph{Both panels:} The highlighted `phase II' is delimited by the turns and corresponds to the field trajectory being aligned with the $\chi$ direction; during this time, $\epsilon_\chi\gg \epsilon_\varphi$. 
   }
    \label{fig:PBHB-background}
\end{figure}

\begin{figure}[h]
    \centering
    \includegraphics[width=\linewidth]{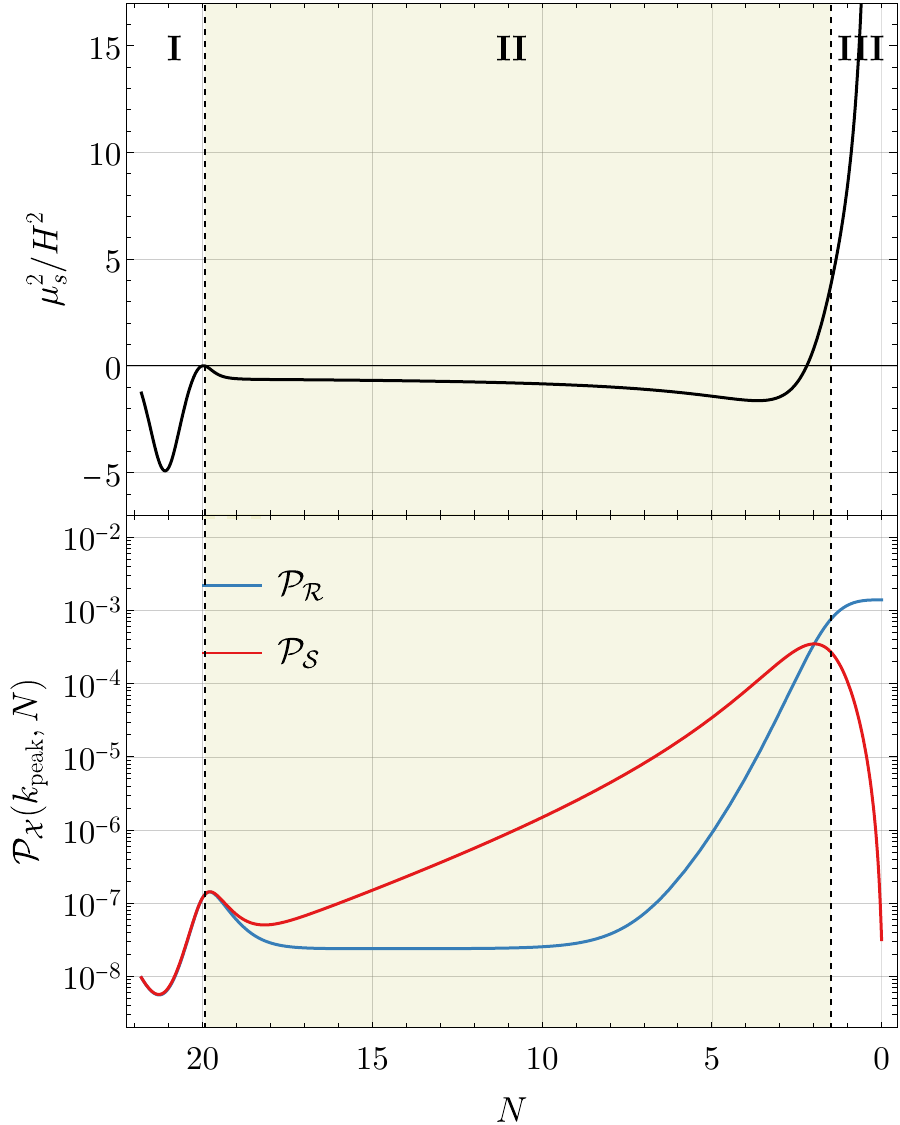}
    \caption{Evolution of the perturbations for the PBH$B$ model of Eq.~\eqref{eq:VPBH-B}, with parameters reported in Table~\ref{tab:PBHB-params} (model `$\chi$-PBH$B$-var').
    \textit{Top panel:} Isocurvature effective mass $\tilde{\mu}_s^2$; the tachyonic instability is highlighted in phase II. 
    \textit{Bottom panel:} Spectra of curvature (${\cal R}$) and isocurvature (${\cal S}$) perturbations 
    for a mode $k_\text{peak}$ that exits the Hubble radius $\sim 21$ e-folds before the end of inflation. The isocurvature mode undergoes tachyonic growth during phase II before transferring power to the curvature perturbation while $\omega \neq 0$ at the transition to phase III. 
    }
    \label{fig:PBHB-perturbations}
\end{figure}

The model parameters used in the figures throughout our main paper are reported in Table~\ref{tab:PBHA-params} for the PBH$A$ model of Eq.~\eqref{eq:VPBH-A} and in Table~\ref{tab:PBHB-params} for the PBH$B$ model of Eq.~\eqref{eq:VPBH-B}. The evolution of the background quantities $\epsilon$ and $\omega$, and of the perturbations ${\cal R}_k$ and ${\cal S}_k$, are shown for model PBH$B$ in Figs.~\ref{fig:PBHB-background} and~\ref{fig:PBHB-perturbations}; these correspond to the evolution of those quantities depicted in Figs.~\ref{fig:PBHA-background} and~\ref{fig:PBHA-perturbations} for the PBH$A$ model. Additional details can be found in our long-form companion paper, Ref.~\cite{Lorenzoni:2025kwn}.

The CMB observables calculated for each of the PBH$A$ and PBH$B$ models considered throughout are summarized in Tables~\ref{tab:PBHA-obs} and~\ref{tab:PBHB-obs}, respectively. Their definitions are given below, and their best-fit observational values are reported in Table~\ref{tab:CMB-obs}.

The time at which the comoving CMB pivot scale, $k_{\rm CMB}=0.05 \text{ Mpc}^{-1}$, first crosses the Hubble radius is found through the standard relation~\cite{Dodelson:2003vq,Liddle:2003as}
\begin{align}
    N_{\rm CMB} &\simeq 62 + \frac{1}{4} {\rm ln} \left( \frac{ \rho_{\rm CMB}^2}{3 \mpl^6 H^2_{\rm e} } \right) + \frac{1-3 w_{\rm reh}}{12(1+ w_{\rm reh})}   \ln\left(\frac{\rho_{\rm rad}}{\rho_{\rm e}}\right) \nonumber \\
    &\simeq 56 \pm 5 \, ,
\end{align}
where $\rho_{\rm CMB}$ is the energy density of the field system at the time the comoving CMB pivot scale exits the Hubble radius during inflation, $\rho_{\rm rad}$ is the energy density when the Universe 
achieves radiation-dominated evolution after inflation and reheating~(see,  e.g., Refs.~\cite{Amin:2014eta,Cook:2015vqa,Martin:2016oyk,Allahverdi:2020bys}), $H_{\rm e}$ and $\rho_{\rm e}$ are evaluated at the end of inflation, and $\omega_{\rm reh}\in \{ -1/3, +1 \}$ is the equation of state during reheating. (The range $\pm 5$ reflects uncertainty in the duration of reheating.) The central value corresponds to instant reheating, with $\omega_{\rm reh}=1/3$.

Experimental efforts to measure anisotropies in the CMB parametrize the primordial power spectrum of curvature perturbations $\prk(k)$ as~\cite{Planck:2015sxf} 
\begin{equation}
    \prk(k)\equiv A_{\rm s} \left(\frac{k}{k_\text{CMB}}\right)^{n_{\rm s}-1+\frac{1}{2}\alpha_{\rm s}\ln\left(\frac{k}{k_\text{CMB}}\right)}\,,
\end{equation}
where $A_{\rm s}=\prk(k_\text{CMB})$ is the magnitude of the power spectrum for modes with comoving wavenumber equal to the CMB pivot scale $k_{\rm CMB}$, $n_{\rm s}$ is the scalar spectral index (i.e. the departure of the slope of $\prk$ from a scale-free spectrum), and $\alpha_{\rm s}$ is the running of the scalar spectral index. The latter are defined as
\begin{align}
    n_{\rm s}&\equiv 1+\left.\frac{\dd\,\ln\prk(k)}{\dd\,\ln k}\right|_{k=k_\text{CMB}}\,,\\
    \alpha_{\rm s}&\equiv\left.\frac{\dd\,n_{\rm s}}{\dd\,\ln k}\right|_{k=k_\text{CMB}}\,.
\end{align}
The tensor-to-scalar ratio is defined as the ratio between the amplitudes of tensor and scalar perturbations at the CMB pivot scale, $r\equiv A_t/A_s$. Here, $A_t\equiv\calP_t(k_\text{CMB})$, the power spectrum of the tensor modes $h_k$. 

The primordial isocurvature fraction is defined as~\cite{Planck:2018jri}
\begin{equation}
    \beta_\text{iso}(k)\equiv\frac{{\cal P}_{\cal S}(k)}{\prk(k)+{\cal P}_{\cal S}(k)}\,,
\end{equation}
and is expected to be small in these models, since isocurvature modes ${\cal S}_k$ decay at the end of the inflationary evolution, when $\tilde{\mu}_s>0$ (see bottom panels of Figs.~\ref{fig:PBHA-perturbations} and~\ref{fig:PBHB-perturbations}). We evaluate $\beta_\text{iso}$ at the CMB pivot scale, corresponding to the scale $k_\text{mid}$ in Ref.~\cite{Planck:2018jri}.

The non-Gaussianity of the model is calculated using the publicly available Python code \texttt{PyTransport}~\cite{Dias:2016rjq,Mulryne:2016mzv}. We find that non-Gaussianity is $\ll{\cal O}(1)$ for both the single-field and spectator models, and that the bispectrum is mildly peaked in the local configuration.
Note that the modes $k_\text{PBH}$ never contribute to this measure since they correspond to much smaller scales than those probed by CMB data, even in the squeezed limit. Planck 2018 CMB  data probes multipoles $\ell$ and hence $k\approx \ell/\eta_0$  \cite{Lyth:2009imm}, where $\eta_0 \simeq 14000$ Mpc is the particle horizon, over a range of ${\cal O}(10^{-3})$, corresponding to $\approx 7$ e-folds of inflation. This in turn restricts the range of $k$ that can be measured in the CMB (including non-Gaussianity) to $k\ll k_\text{PBH}$. 
We quote the values of $f_{\rm NL}^{\rm local}$ in Tables~\ref{tab:PBHA-obs} and~\ref{tab:PBHB-obs}, which were found to be slightly larger than $f_{\rm NL}^{\rm equil}$ and $f_{\rm NL}^{\rm ortho}$.

Primordial black holes form from the gravitational collapse of overdensities after the end of inflation. In order for PBHs to form, the dimensionless power spectrum must exceed the threshold ${\cal P}_{\cal R} (k_{\rm peak}) \geq 10^{-3}$ \cite{Young:2019yug,Kehagias:2019eil,Escriva:2019phb,DeLuca:2020ioi,Musco:2020jjb,Escriva:2021aeh}, where $k_{\rm peak}$ is the comoving wavenumber at which ${\cal P}_{\cal R} (k)$ reaches its maximum value. The resulting population of PBHs forms with a mass distribution whose peak depends on $k_{\rm PBH} \simeq k_{\rm peak}$. 
The characteristic mass is proportional to the mass contained within a Hubble radius at the time of collapse, $M_{\rm PBH,f}=\gamma M_H(t_{\rm f})$, where $\gamma\simeq0.2$ is an efficiency factor~\cite{Carr:1975qj}, the subscript $\rm f$ refers to the time of formation, and $M_H(t)\equiv 4\pi \rho(t)/(3 H^3(t))$. This can be recast as~\cite{Ozsoy:2023ryl}
\begin{equation}
    \frac{M_{\rm PBH,f}}{30\, M_{\odot}} \simeq \left(\frac{\gamma}{0.2}\right)\left(\frac{g_{*}\left(T_{\rm f}\right)}{106.75}\right)^{-\frac{1}{6}}\left(\frac{k_{\rm PBH}}{3.2 \times 10^{5}\, \mathrm{Mpc}^{-1}}\right)^{-2}\,.
\end{equation}
In this work, we take $\gamma=0.2$, consider the effective number of relativistic degrees of freedom to be $g_*(T_{\rm f})=106.75$, and we translate this mass into grams using $M_\odot\simeq2.0\times10^{33}~{\rm g}$.

\begin{table*}[h]
    \centering
    \begin{tabular}{@{\extracolsep{10pt}}c|*{8}c}
        \hline\hline
        Model 
        & $V_0~[\mpl^4]$ & $\varphi_d~[\mpl]$ 
        & $m_{\chi}~[\mpl]$ 
        & $\chi_i~[\mpl]$
        \\ 
        \hline % ----- PARAMETER SETS BELOW ----- %
        PBH$A$-fid 
        & $7.7\times 10^{-11}$ & $2.10$ 
        & $-$ & $-$
        \\
        PBH$A$-var
        & $6.8\times 10^{-11}$ &
        $2.10\times(1-4\times 10^{-4})$
        & $-$ & $-$
        \\
        $\chi$-PBH$A$-fid
        & $7.7\times 10^{-11}$ & $2.10$ 
        & $1\times10^{-7}$ & $1$
        \\
        $\chi$-PBH$A$-var
        & $9.7\times 10^{-11}$ &
        $2.10\times(1-4\times 10^{-4})$
        & $8\times10^{-7}$ & $3.3$
        \\
        \hline\hline
    \end{tabular}
    \caption{Values of the parameters considered for the PBH$A$ model. For each case, we fix the values $M=1/2$, $A=1.421\times 10^{-3}$, and $\sigma=1.7138\times10^{-2}$ (in units of $\mpl$), and always consider the initial value $\varphi_i=3.5~\mpl$ for the inflaton. (Inflationary dynamics for these models are largely independent of $\varphi_i$, as long as it allows for sufficient e-folds of inflation.) We take $\dot{\varphi}_i = \dot{\chi}_i = 0$; when using PyTransport, we set $\dot{\varphi}_i = 10^{-10}$ to avoid numerical issues. The reference model PBH$A$-fid uses similar parameters to the ones considered in Ref.~\cite{Cole:2023wyx}, while PBH$A$-var shifts the parameter $\varphi_d\rightarrow\varphi_{d,\text{fid}}\times(1-4\times 10^{-4})$. The corresponding models $\chi$-PBH$A$-fid and $\chi$-PBH$A$-var include the effect of a spectator field.
    }
    \label{tab:PBHA-params}
\end{table*}

\begin{table*}[h]
    \centering
    \makebox[\textwidth][c]{
    \begin{tabular}{@{\extracolsep{10pt}}c|*{7}c|*{2}c}
        \hline\hline
        Model
        & $A_s$ & $n_s$ & $\alpha_s$ & $r$ & $\beta_{\rm iso}$ & $f_{\rm NL}^{\rm local}$ & $M_{\rm PBH}~
        [{\rm g}]$
        & $V_\text{S}/V_\text{PBH}$ & $m_\chi/H$
        \\ 
        \hline % ----- PARAMETER SETS BELOW ----- %
        PBH$A$-fid  
        & $2.09\times 10^{-9}$ & $0.9661$ & $-7.6\times 10^{-4}$ & $0.002$ & $-$ & $0.02$ & $5.0 \times 10^{18}$
        & $-$ & $-$
        \\
        PBH$A$-var 
        & $2.10\times 10^{-9}$ & $0.9689$ & $-6.4\times 10^{-4}$ & $0.002$ & $-$ & $0.02$ & $-$
        & $-$ & $-$
        \\
        $\chi$-PBH$A$-fid 
        & $2.09\times 10^{-9}$ & $0.9661$ & $-7.6\times 10^{-4}$ & $0.002$ & $1.8\times 10^{-4}$ & $0.02$ & $5.6 \times 10^{18}$
        & $7\times10^{-5}$ & $0.02$
        \\
        $\chi$-PBH$A$-var 
        & $2.11\times 10^{-9}$ & $0.9670$ & $-6.9\times 10^{-4}$ & $0.003$ & $1.4\times10^{-4}$ & $0.02$ & $5.7 \times 10^{18}$
        & $0.02$ & $0.14$
        \\
        \hline\hline
    \end{tabular}
    }
    \caption{Observables for all the parameter sets considered for the PBH$A$ model (as defined in Table~\ref{tab:PBHA-params}). CMB scales exit at $N_\star=56$. The non-Gaussianities are mildly peaked in the local configuration, and $\beta_\text{iso}$ is taken at the scale $k=0.05~\text{Mpc}^{-1}$ (i.e. $k_\text{mid}$ in Ref~\cite{Planck:2018jri}). The `spectator-ness' measures $V_\text{S}/V_\text{PBH}$ and $m_\chi/H$ are calculated at the time $t_\text{CMB}$ when CMB scales first exit the Hubble radius.
    }
    \label{tab:PBHA-obs}
\end{table*}

\begin{table*}[h]
    \centering
    \begin{tabular}{@{\extracolsep{10pt}}c|*{7}c}
        \hline\hline
        Model 
        & $\lambda$ & $v~[\mpl]$ 
        & $m_{\chi}~[\mpl]$ 
        & $\chi_{i}~[\mpl]$
        \\ 
        \hline % ----- PARAMETER SETS BELOW ----- %
        PBH$B$-fid  
        & $1.16\times 10^{-6}$ & $0.19669$ 
        & $-$ & $-$ 
        \\
        PBH$B$-var
        & $9.10\times 10^{-7}$ &
        $0.19669\times (1-4\times10^{-3})$
        & $-$ & $-$ 
        \\
        $\chi$-PBH$B$-fid  
        & $1.16\times 10^{-6}$ & $0.19669$
        & $1\times10^{-7}$ & $1$ 
        \\
        $\chi$-PBH$B$-var  
        & $1.19\times 10^{-6}$ &
        $0.19669\times (1-4\times10^{-3})$
        & $7\times10^{-7}$ & $2.5$ 
        \\
        \hline\hline
    \end{tabular}
    \caption{Values of the parameters considered for the PBH$B$ model. For each case, we fix the values $a=0.719527$ and $b=1.500016$, and always consider the initial value $\varphi_i=3~\mpl$ for the inflaton. (Inflationary dynamics in these models are largely independent of $\varphi_i$, as long as it allows for sufficient e-folds of inflation.) We take $\dot{\varphi}_i = \dot{\chi}_i = 0$; when using PyTransport, we set $\dot{\varphi}_i = 10^{-10}$ to avoid numerical issues.) 
    The reference model PBH$B$-fid reproduces the results introduced in Ref.~\cite{Garcia-Bellido:2017mdw}, modified to be in agreement with the latest CMB constraints and producing PBHs in the asteroid-mass range, while PBH$B$-var shifts the parameter $v\rightarrow v_{\text{fid}}\times(1-4\times10^{-3})$. The corresponding models $\chi$-PBH$B$-fid and $\chi$-PBH$B$-var include the effect of a spectator field. 
    }
    \label{tab:PBHB-params}
\end{table*}

\begin{table*}[h]
    \centering
    \makebox[\textwidth][c]{
    \begin{tabular}{@{\extracolsep{10pt}}c|*{7}c|*{2}c}
        \hline\hline
        Model 
        & $A_s$ & $n_s$ & $\alpha_s$ & $r$ & $\beta_{\rm iso}$ & $f_{\rm NL}^{\rm local}$ & $M_{\rm PBH}~[{\rm g}]$
        & $R_V$ & $m_\chi/H$
        \\ 
        \hline % ----- PARAMETER SETS BELOW ----- %
        PBH$B$-fid  
        & $2.10\times 10^{-9}$ & $0.9599$ & $-1.8\times 10^{-3}$ & $0.005$ & $-$ & $0.02$ & $1.1 \times 10^{17}$
        & $-$ & $-$
        \\
        PBH$B$-var 
        & $2.10\times 10^{-9}$ & $0.9665$ & $-8.2\times 10^{-4}$ & $0.004$ & $-$ & $0.01$ & $-$ 
        & $-$ & $-$
        \\
        $\chi$-PBH$B$-fid 
        & $2.09\times 10^{-9}$ & $0.9595$ & $-1.1\times 10^{-3}$ & $0.005$ & $3.6\times10^{-4}$ & $0.02$ & $1.2 \times 10^{17}$
        & $3\times10^{-5}$ & $0.01$
        \\
        $\chi$-PBH$B$-var 
        & $2.09\times 10^{-9}$ & $0.9640$ & $-1.0\times 10^{-3}$ & $0.005$ & $2.9\times10^{-4}$ & $0.02$ & $4.0 \times 10^{16}$
        & $0.008$ & $0.10$
        \\
        \hline\hline
    \end{tabular}
    }
    \caption{Observables for all the parameter sets considered for the PBH$B$ model (as defined in Table~\ref{tab:PBHB-params}). CMB scales exit at $N_\star=56$. The non-Gaussianities are mildly peaked in the local configuration, and $\beta_\text{iso}$ is taken at the scale $k=0.05~\text{Mpc}^{-1}$ (i.e. $k_\text{mid}$ in Ref~\cite{Planck:2018jri}). The `spectator-ness' measures $V_\text{S}/V_\text{PBH}$ and $m_\chi/H$ are calculated at the time $t_\text{CMB}$ when CMB scales first exit the Hubble radius.}
    \label{tab:PBHB-obs}
\end{table*}

\begin{table*}[h]
    \centering
    \makebox[\textwidth][c]{
    \begin{tabular}{@{\extracolsep{10pt}}*{7}c}
        \hline\hline
        $A_{\rm s}~[\times10^{-9}]$ & $n_{\rm s}$ & $\alpha_{\rm s}$ & $r$ & $\beta_{\rm iso}$ & $f_{\rm NL}^{\rm local}$ & $M_{\rm PBH}~[{\rm g}]$
        \\ 
        \hline % ----- BEST-FIT VALUES ----- %
        $2.10\pm0.03$ % A_s
        & $0.9665\pm0.0038$ % n_s
        & $-0.0045\pm0.0067$ % alpha_s
        & $<0.037$ % r
        & $<0.001$ % beta_iso
        & $-0.9 \pm 5.1$ % f_NL^local
        & $[10^{17} - 10^{23}]$ % M_PBH
        \\
        \hline\hline
    \end{tabular}
    }
    \caption{Best-fit value for CMB observables from Planck 2018 CMB data~\cite{Planck:2019kim,Planck:2018vyg,Planck:2018jri,BICEP:2021xfz} and PBH constraints~\cite{Khlopov:2008qy,Carr:2009jm,Sasaki:2018dmp,Carr:2020gox,Carr:2020xqk,Green:2020jor,Escriva:2021aeh,Villanueva-Domingo:2021spv,Escriva:2022duf,Gorton:2024cdm}. 
    }
    \label{tab:CMB-obs}
\end{table*}

%%%%%%%%%%%%%%%%%%%%%%%%%%%%%%%%%%%
% ---------------------------------
\end{document}